\documentclass[]{spie}  
\usepackage[]{graphicx}
\usepackage{amssymb}
\usepackage[table]{xcolor}   
\usepackage[width=.85\textwidth,font=small]{caption}
\usepackage{longtable}
\usepackage[pdftex,
 pdfauthor={Brendan Crill et al.},
 pdftitle={Space Technology for Directly Imaging and Characterizing Exo-Earths},
 pdfdisplaydoctitle=True]{hyperref}

\usepackage{array}
\newcolumntype{L}[1]{>{\raggedright\let\newline\\\arraybackslash\hspace{0pt}}m{#1}}
\newcolumntype{K}[1]{>{\raggedright\centering\arraybackslash}m{#1}}

\title{Space Technology for Directly Imaging and Characterizing Exo-Earths} 

\author{Brendan P. Crill\supit{a} and Nicholas Siegler\supit{a}
\skiplinehalf
\supit{a}Jet Propulsion Laboratory, California Institute of Technology, 4800 Oak Grove, Pasadena CA 91109, USA
}

\authorinfo{Further author information:\\B.P.C. E-mail: {\tt bcrill@jpl.nasa.gov}, Telephone: +1 818 354 5416}

\begin{document} 
  \maketitle 
  \begin{abstract}
  The detection of Earth-like exoplanets in the habitable zone of their stars, and their spectroscopic characterization in a search for biosignatures, requires starlight suppression that exceeds the current best ground-based performance by orders of magnitude.   
  The required planet/star brightness ratio of order 10$^{-10}$ at visible wavelengths can be obtained by blocking stellar photons with an occulter, either externally (a starshade) or internally (a coronagraph) to the telescope system, and managing diffracted starlight, so as to directly image the exoplanet in reflected starlight.
  Coronagraph instruments require advancement in telescope aperture (either monolithic or segmented), aperture obscurations (obscured by secondary mirror and its support struts), and wavefront error sensitivity (e.g. line-of-sight jitter, telescope vibration, polarization). 
The starshade, which has never been used in a science application, benefits a mission by being decoupled from the telescope, allowing a loosening of telescope stability requirements.  In doing so, it transfers the difficult technology from the telescope system to a large deployable structure (tens of meters to greater than 100~m in diameter) that must be positioned precisely at a distance of tens of thousands of kilometers from the telescope.
    We describe in this paper a roadmap to achieving the technological capability to search for biosignatures on an Earth-like exoplanet from a future space telescope. Two of these studies, HabEx and LUVOIR, include the direct imaging of Earth-sized habitable exoplanets as a central science theme.
  
  \end{abstract}

  
\section{INTRODUCTION}
\label{sec:intro} 
 
The capability to directly image and spectrally characterize exoplanets in the habitable zone of their stars is a key step in the search for life in the Universe.   The 2010 Astronomy and Astrophysics Decadal Survey \textit{New Worlds, New Horizons}\cite{Decadal2010} and NASA's response in the Astrophysics Implementation Plan \cite{AIP2016} placed a high priority on developing technology for a space mission to achieve this goal, a task managed by NASA's Exoplanet Exploration Program (ExEP).   In anticipation that the upcoming 2020 Decadal Survey could recommend that NASA develop a mission capable of searching for biosignatures on Earth-like planets\footnote{We focus here on Earth-like planets orbiting Sun-like stars (FGK stars); Earth-sized exoplanets orbiting M dwarf stars are likely to be observable by ground-based observatories and will therefore be studied on a different timeline.} with direct imaging, potentially following the Wide-Field Infrared Space Telescope (WFIRST) as an astrophysics mission, technologies must be advanced to a level of maturity that allows a timely development of the selected mission. 

\begin{figure}
\begin{center}
\includegraphics[width=\columnwidth]{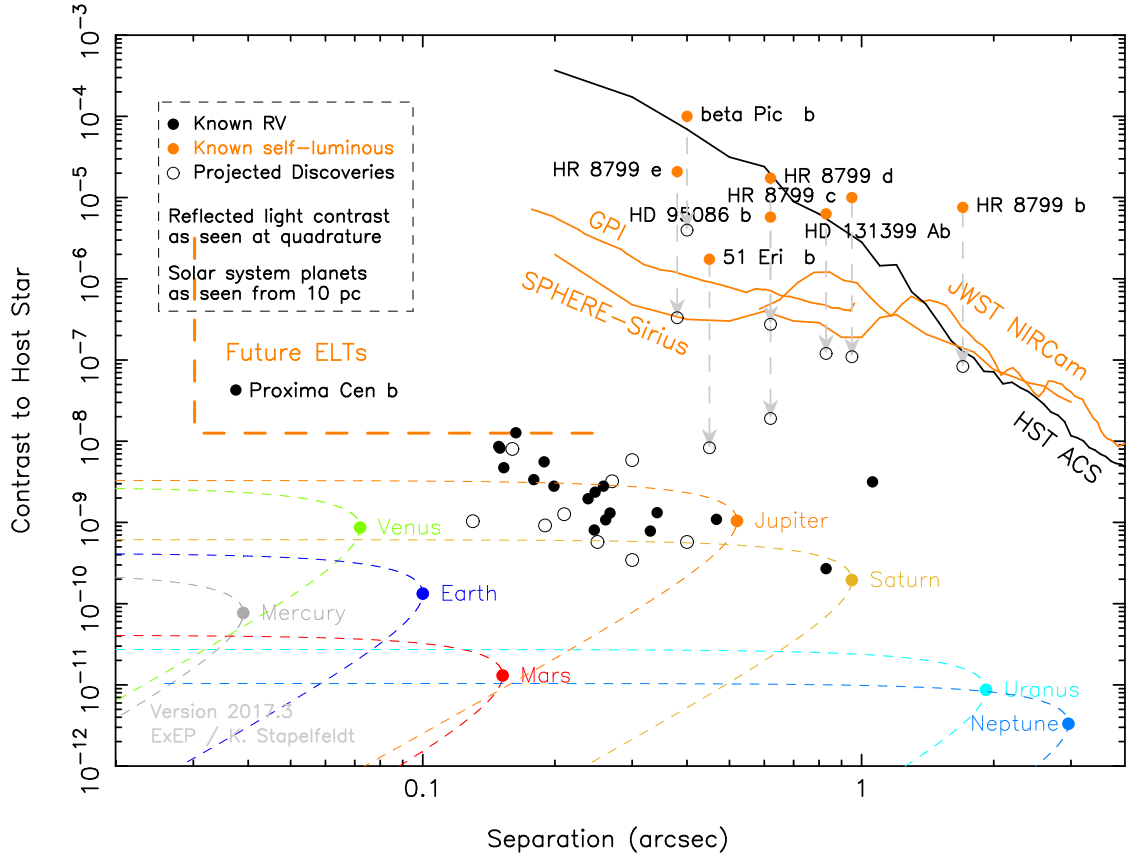}
\caption{\label{fig:ContrastPlot} \textbf{Exoplanet contrast vs. angular separation\cite{TechnologyPlanAppendix2017}.}  Contrast (ratio of planet brightness to host star brightness) versus angular separation. The filled orange circles indicate the direct imaging of young, self-luminous planets imaged in the near-infrared by ground-based telescopes. Contrasts for the planets of the Solar System are for analogous planets placed 10~pc away. The solid black dots are contrast estimates of measured radial velocity planets.  The solid orange curves show measured performance of ground-based coronagraphs: the GPI curve shows typical performance, while the SPHERE curve shows the best achieved performance to-date on Sirius. Achieved performance with HST/ACS coronagraphic masks, and the predicted performance of JWST/NIRCam masks are also shown. The dashed orange curve assumes future ELTs will reach 10$^{-8}$ contrast at 0.03~arcsec\cite{stapelfeldt2005}.   Planets discovered in the near-infrared are shown with vertical arrows pointing to the predicted contrast ratios at visible wavelengths.}
\end{center}
\end{figure}

Successful direct imaging of an Earth-like planet at 10~pc requires improvement in the combination of high angular resolution and several orders of magnitude improvement in starlight suppression to achieve the required contrast: the brightness ratio of reflected starlight at visible wavelengths from an Earth-like exoplanet orbiting a Sun-like star is of order 10$^{-10}$ (See  Fig.~\ref{fig:ContrastPlot}).   Advancements in adaptive optics technology and the construction of 30~m class telescopes (GMT, TMT, E-ELT) will enable deep contrasts at very high angular resolution, but the use of a coronagraph on the ground is limited by the ability to correct atmospheric turbulence with adaptive optics to roughly 10$^{-8}$.  Hence, direct imaging of exo-Earths will require a space telescope\cite{stapelfeldt2005}.  To achieve these ambitious goals in space, a number of technologies must be advanced. 


WFIRST, scheduled to launch in the mid-2020's, will provide an essential technology demonstration with a coronagraph instrument (CGI) that uses a deformable mirror to correct wavefront aberrations, the first to do so in space.  Still in Phase A, the WFIRST mission has recently matured its coronagraph to TRL~5.  The CGI is designed to achieve contrast sensitivities between 10$^{-8}$ and 10$^{-9}$ and to be the first instrument to directly image mature gas and ice giants like our own Jupiter and Neptune. This will be an important step across the gap from today's capability to the needs of a future mission (See Fig.~\ref{fig:ContrastPlot}).

Additionally, at this time, the WFIRST project is studying the benefits and impacts of being compatible with a potential future starshade mission.  While NASA does not yet have plans to initiate a starshade flight project, NASA is studying the scientific potential, cost, and risks of starshade-based observations with WFIRST.  The recommendations of the 2020 Decadal Survey will guide NASA's decision on whether to initiate a starshade project. 

The Habitable Exoplanet Imaging Mission (HabEx), and Large Ultraviolet-Optical-Infrared Surveyor (LUVOIR) mission concepts share Earth-like exoplanet characterization as a central science theme.  These missions are being designed around the direct imaging of exoplanets and the telescope/spacecraft systems are being designed in concert with a coronagraph and/or starshade.  These two mission concepts encapsulate the current state of the science interests and technical capabilities of the exoplanet science and technology communities, providing the 2020 Decadal Survey committee with options for a future life-finding mission.  The technology needs of these missions, which are still under study, can be used as a target to help guide technology development investments in the coming years. 

The Origins Space Telescope (OST) mission concept includes a coronagraph operating in the mid-infrared that could observe and characterize exoplanets in thermal emission, where planet/star brightness ratios are more favorable.  The technology needs for the mid-infrared coronagraph overlap somewhat with those of LUVOIR and HabEx operating at shorter wavelengths.  Due to angular resolution constraints at longer wavelengths, the OST direct imaging science case focuses on imaging and characterizing atmospheres of warm and cool gas giants, and perhaps ice giants.  OST's main contribution to exoplanet science will be in transmission spectroscopy.

The ExEP annually revisits the technology needs of future exoplanet missions and invites the community to review and offer additional needed technologies and capabilities.  These are summarized in the ExEP Technology Plan Appendix\cite{TechnologyPlanAppendix2017}, a primary source for this work.  These technology needs and their development have also been described in other recent publications\cite{morgan2015,coulter2016}.  

In this paper, we review the technology development path for a space mission capable of discovering and characterizing exo-Earths.  Technologies needed to achieve exoplanet science goals are described in Sect.~\ref{sec:technology}.  In Sect.~\ref{sec:path_forward}, we look at the path ahead and survey the current status of technology development, including how far currently planned missions will take us, and describe additional steps that are needed.

  \section{TECHNOLOGY FOR EXO-EARTH IMAGING}
  \label{sec:technology}

 \subsection{KEY MISSION PARAMETERS}

A mission aiming to discover and characterize Earth-like exoplanets in the habitable zones of their stars must have appropriate capabilities in angular resolution, contrast (starlight suppression), contrast stability, and detection sensitivity.  The exact requirements in each of these areas depend on the approach taken by a given mission, but in most cases, the ambitious science goals require unprecedented levels of performance.  The trade studies performed by recent and ongoing concept studies, determine mission parameters that maximize the yield of detected and characterize exoplanets in the habitable zone of their stars\cite{stark2016} and place requirements on technology for a mission.  The Exo-S\cite{exos2015} and Exo-C\cite{exoc2015} studies were centered around direct imaging, as are the HabEx and LUVOIR studies.

Traub and Oppenheimer (2010) \cite{Traub2010} reviewed a number of key concepts in exoplanet direct imaging, for a sense of generic mission parameters.  Surveying habitable zones of Sun-like stars in a greater volume of space and detecting planets through diffuse exo-zodiacal dust drives one to large (at least 4~m diameter)  primary mirrors\cite{stark2015}.  Larger telescopes also allow spectra at longer wavelength of an exoplanetary system as $\lambda/D$ increases; many methane, carbon dioxide, and water features can be seen longward of 1~$\mu$m.\cite{turnbull2006}  The brightness ratio of an Earth-like planet to a Sun-like star seen in reflected visible light is of order $10^{-10}$, requiring unprecedented starlight suppression, either with a high-contrast coronagraph (using active wavefront control technology) or a starshade.   Nulling interferometry as described in the Terrestrial Planet Finder concept\cite{TPFI} is a third approach to starlight suppression; its technology development is further behind the internal and external occulters.  The low absolute brightness of exoplanets (the V-band absolute magnitude of Earth is roughly 30) places requirements both on detection sensitivity and on the stability of the telescope and starlight suppression system.  Typically, wavefront stability must be better than 1~nanometer on the timescale of a detector exposure, depending on the sensitivity of a particular coronagraph design to different spatial modes of wavefront error. 

Table~\ref{tbl:TechnologyGaps} summarizes the technology areas that must be advanced to image exo-Earths, listing the needed capabilities, the current state-of-the-art, and an estimate of predicted level of performance in 2020 due to ongoing activities in these areas.  The HabEx and LUVOIR studies represent two different approaches to exo-Earth study and are defining the technology needed for flight.   LUVOIR's primary architecture is currently a 15~m segmented mirror, using a coronagraph for starlight suppression.  HabEx is planning an off-axis 4~m monolith primary mirror and both a starshade and a coronagraph.   
 \scriptsize
 \begin{center}
 \rowcolors{2}{white}{gray!10}
 \begin{longtable}{L{3cm} L{4.3cm} L{4.3cm} L{4.3cm}}
 \hline
 Technology & Current State-of-the-Art & Projected 2020 Capability& Needed for Flight\\
 \hline
 \endhead
Ultra-stable Opto-mechanical System and  Large Aperture Monolith Mirror&
3.5m sintered SiC with $<$ 3 $\mu$m SFE (Herschel); 2.4m ULE with $\sim$10 nm SFE (HST)
 Lightweighting waterjet cutting to 14" for flight, $>$18"  and fused core on lab sub-scale demos.& 
HabEx systems-level design will set requirements at a sub-system and component level.  Missing capabilities will be better understood.& 
HabEx 4 m monolith: SFE $<$ 10 nm rms, Mirror stiffness and thermal stability for wavefront stability better than 100 pm rms on timescales of minutes.    \\
  
Ultra-stable Opto-mechanical System and Large Aperture Segmented Mirror&
Mirror Surface Figure Error: $<$ 7 nm RMS (JWST Be); Wavefront Error Stability:$\sim$50 nm / control step.  6 DOF, 1-m class SiC and ULE, $<$20 nm SFE, and $<$5 nm wavefront stability over 4 hr with thermal control & 
Systems-level study (LUVOIR, etc.) will set requirements at a sub-system and component level. Missing capabilities will be better understood.& 
Systems-level architecture studies and error budgets flow this top level requirement down to component-level performances.  Likely budgets: Mirror Surface Figure Error: $<$ 5 nm RMS; Wavefront Error Stability:$<$10 pm / control step.      
High-speed ($>$1 Hz) closed-loop control of segment rigid body motions to $\sim$10~pm. Control-loop bandwidths 100~Hz.  Low-noise, stable electronics for control.
 \\
 
 Clear-aperture Coronagraph & 
 6$\times$10$^{-10}$ raw contrast, 10\% bandwidth across angles of 3-16 $\lambda$/D demonstrated with a linear mask and an unobscured pupil in a static vacuum lab env't (Hybrid Lyot).   Latest HLC design is resilient to tip/tilt, astigmatism.  High-charge vector vortex designs have resilience against higher order spatial modes of wavefront errors.
 &
 Decadal Survey Testbed Phase I will demonstrate 10$^{-10}$ contrast, 10\% bandwidth in a static vacuum lab environment, inner working angle 3$\lambda/D$.&  
 Coronagraph masks and optics capable of creating circularly symmetric dark regions in the focal plane enabling raw contrasts $\le$10$^{-10}$, with minimal degradation from polarization aberration, IWA $\le$3 $\lambda$/D, throughput $\ge$10\%, and bandwidth $\ge$ 10\% on obscured/ segmented pupils in a simulated dynamic vacuum lab environment. 
 \\
 
 Segmented-aperture Coronagraph &
WFIRST CGI has demonstrated 2$\times$10$^{-9}$ at 10\% BW for an obscured, monolith aperture. Models for 10$^{-10}$ performance developed without robustness. Demonstrations with segmented apertures with VNC achieved 5$\times$10$^{-9}$ narrow band. & 
 Decadal Survey Testbed Phase II will demonstrate 10$^{-10}$ contrast, 10\% bandwidth on a simulated static segmented mirror in a static vacuum lab environment, inner working angle 3$\lambda/D$.& 
Raw contrast of 10$^{-10}$ at IWA of $\sim$3$\lambda$/D in the visible band, 10\% band, segmented/obscured aperture. \\
 
 Ultra-low Noise Near-infrared (0.9 -- 2.5~$\mu$m) Detectors & 
 HgCdTe photodiode: read noise $\sim$2~e$^{-}$ rms with multiple non-destructive reads; dark current $<$0.001~e$^{-}$/s/pix; very radiation tolerant (JWST). 
HgCdTe APDs: dark current $\sim$10-20~e$^{-}$/s/pix, read noise $<<$1 e$^{-}$ rms, and $<$1k$\times$1k format.
Cryogenic (superconducting) MKID or TES detectors have zero read noise and dark current; radiation tolerance is unknown.
 & 
 Improvement in HgCdTe APD dark current; further experience with HgCdTe photoarray performance on-orbit (JWST, Euclid)
 & 
 $<$10$^{-3}$~e$^{-}$/px/s dark current; $<$1~e$^{-}$ read noise; $\ge$2k$\times$2k format. \\
 
 Ultra-low Noise Visible-light (200--1000~nm) Detectors & 
 1k$\times$1k WFIRST EMCCD detectors provide dark current of 7$\times$10$^{-4}$~e$^{-}$/px/s; CIC 2.3$\times$10$^{-3}$~e$^{-}$/px/fram; effective read noise $<$0.2~e$^{-}$ RMS after irradiation at 165~K. & 
rad hardness improvement (WFIRST) & 
 $<$10$^{-4}$~e$^{-}$/px/s dark current; $<$0.1 e$^{-}$ read noise; $<$ 3$\times$10$^{-3}$ CIC; format$\ge$2k$\times$2k .\\
 
 Starshade Diffracted Starlight Suppression & 
 Validated optical model with demonstrated 10$^{-6}$ suppression at white light, 58~cm mask, and F$_{1.0}$ =210; 3$\times$10$^{-8}$ suppression demonstrated at F$_{1.0}$ =15;
1.3$\times$10$^{-7}$ suppression demonstrated at F$_{1.0} \approx$ 50& 
demonstrated $\le$10$^{Ð8}$ in scaled flight-like geometry (F$_{1.0}$=15)& 
 Experimentally validated models with total starlight suppression $\le$10$^{Ð8}$ in scaled flight-like geometry, with 5$<$F$_{1.0}<$40 across a broadband optical bandpass. Validated models are traceable to 10$^{-10}$ contrast system performance in space.\\ 
 
 Starshade Scattered Sunlight Suppression & 
 Machined graphite edges meet all specs but edge radius-of-curvature ($\le$10~$\mu$m); etched metal edges meet all specs but in-plane shape tolerance for long segments.&
1~m edge segments demonstrated to meet specifications  & 
  Integrated petal optical edges maintaining precision in-plane shape requirements after deployment trials and limit solar glint contributing $<$10$^{-10}$ contrast at petal edges.\\ 
  
 Starshade Lateral Formation Sensing &
 Performance demonstrations at JPL meet the formation flying sensing needs of a WFIRST-rendezvous starshade mission. A LUVOIR or HabEx mission involves a larger starshade at a further distance, but the sensing needs are similar given the larger telescope aperture and similar angular diameter of the starshade.&
Scaled lab demonstration of formation sensing that feeds a control loop.
  &
   Demonstrate sensing lateral errors $\le$0.20 m accuracy at scaled flight separations (1~mas bearing angle). Control algorithms demonstrated with scaled lateral control errors corresponding to $\le$1~m.\\ 
   
 Starshade Petal Positioning & 
Perimeter Truss Deployment method: Petal shape tolerance ($\le$1~mm) verified with low fidelity 12~m prototype, no optical shield, 6~m protoype demonstrated with optical shield; no environmental testing.& 
Deployment method downselected.
 & 
Shape tolerances demonstrated to $\le$1~mm (in-plane envelope) with flight-like, minimum half-scale structure, simulated petals, opaque structure, and interfaces to launch restraint after exposure to relevant environments.\\ 

Starshade Petal Shape and Stability & 
Manufacturing tolerance ($\le$100~$\mu$m) verified with low fidelity 6~m prototype and no environmental tests. Petal deployment tests conducted but on prototype petals to demonstrate rib actuation; no shape measurements.&
Deployment method downselected.
 &
 Shape tolerances demonstrated to $\le$100~$\mu$m (in-plane envelope) with flight-like, minimum half-scale petal fabricated and maintains shape after multiple deployments from stowed configuration. \\ 
 
 \hline
 \rowcolor{white}
 \caption{\label{tbl:TechnologyGaps} \textbf{Technologies to be developed for exo-Earth direct imaging and characterization.}}
 \end{longtable}
 \end{center}
\normalsize
 
  \subsection{CORONAGRAPH}
 \label{sec:coronagraphs}
 A coronagraph is an optical instrument that enables high-contrast imaging by blocking on-axis light from a bright source\cite{lyot1932}, enabling nearby faint sources to come into view. 
 Modern high contrast coronagraphs designed for exoplanet imaging in the presence of a bright star, incorporate masks or mirrors in pupil and image planes that block starlight, direct diffraction to where it can be removed, and reduce residual scattered starlight by adjusting the amplitude and phase of the incoming starlight. They use adaptive optics, with feedback to fast steering mirrors and deformable mirrors, to improve contrast and contrast stability by correcting static and dynamic wavefront errors. 

Coronagraphic masks and designs have improved dramatically in the past decade.  The Technology Development for Exoplanet Missions (TDEM) portion of NASA's Strategic Astrophysics Technology (SAT) program\cite{tdem_website} has funded advancement in a number of coronagraph designs, including Phase-induced Amplitude Apodization (PIAA), PIAA Complex Mask Coronagraph (PIAACMC), Hybrid Lyot coronagraph (HLC), Visible Nuller (VNC), Vortex Coronagraph (VC), and Shaped Pupil Coronagraph (SPC), including demonstrations in the ExEP's High Contrast Imaging Testbed (HCIT) vacuum test facility.  Most importantly for advancing coronagraph technology, the WFIRST project has baselined CGI which uses HLC\cite{seo2016} and SPC\cite{cady2015} designs, along with a low order wavefront sense and control (LOWFS/C) system\cite{shi2015}, a first for space.   Project-funded technology development towards achieving CGI's milestones has demonstrated better than $10^{-8}$ broadband contrast with the HLC and SPC designs (for working angles 3 to 9 $\lambda$/D) in a dynamical environment (i.e. including expected drifts due to spacecraft thermal effects and pointing jitter), enough to directly image exo-Jupiters and exo-Saturns.  These demonstrations marked CGI's advancement to TRL~5. 

More work is needed to achieve contrast and stability for exo-Earth direct imaging.  The deepest contrast demonstrated to-date in a laboratory environment is 1.2$\times$10$^{-10}$ with a 3 $\lambda$/D inner working angle, using a Hybrid Lyot Coronagraph\cite{trauger2007} in the HCIT, though with narrow (2\%) bandwidth and a linear mask.  One testbed in HCIT, called the Decadal Survey Testbed (DST), is currently being designed to enable laboratory demonstrations of 10\% bandwidth 10$^{-10}$ contrast in anticipation of competed demonstrations to support HabEx and LUVOIR.    The SAT/TDEM program is currently funding development of Vortex coronagraph masks and Visible Nuller Coronagraphs.  Additionally, polarization effects internal to the telescope and coronagraph may limit the contrast performance of some designs and is being investigated under a TDEM award\cite{breckinridge_tdem_wp}.  

Deformable mirrors (DMs) are an essential element for coronagraphy and are being matured with WFIRST's technology program.  A DM typically used internal to a high-contrast coronagraph is a reflective optic of order 10~cm size with an array of actuators in a grid pattern that can deform the reflective surface and thereby correct wavefront phase.  Normally the primary pupil is mapped to the DM, so the actuator count of the DM determines the minimum scale of wavefront errors that can be corrected.  WFIRST uses a DM with a 48$\times$48 grid of actuators. The actuator resolution is defined by how finely the voltage to each actuator can be adjusted (normally $<$1~nm) and the total stroke of each actuator (maximum surface adjustment) is about 0.5~$\mu$m.  Future missions are likely to need larger format DMs (up to 96$\times$96), but mosaicking of the actuation stages is possible.  If greater stroke is needed, however, further development work may be needed.

A coronagraph for a future large space telescope may also need to work with an obscured aperture, and handle the more complicated diffraction due to segment gaps, support obscurations, and an on-axis secondary mirror.  The ExEP launched a Segmented Coronagraph Design Analysis study to investigate this challenge\cite{zimmerman2016,ruane2016}, with the goal of proving that a robust coronagraph design exists for segmented and obscured pupils.  So far, the Apodized Pupil Lyot Coronagraph (APLC) approach\cite{NDiaye2015} has achieved the most success.   

In addition to obtaining a deep contrast, the contrast level must be stable for the long time scales required for a deep integration, on the order of hours or days.  Some disturbances due to telescope jitter, pointing drifts, and lower order wavefront aberrations due to telescope thermal drifts can be sensed and maintained with LOWFS/C internal to the coronagraph.  Additionally, the telescope and spacecraft system must be designed with extreme stability in mind.  The HabEx and LUVOIR studies, in performing system-level trade studies and creating reference designs, will provide important knowledge about the requirements that subsystems must meet for a 10$^{-10}$ coronagraph.
 
 A coronagraph for use in the mid-infrared on OST can build on technologies considered for the Japanese Space Infrared Telescope for Cosmology and Astrophysics (SPICA)\cite{enya2017}.  An OST coronagraph requires similar study of compatibility with obscured/segmented telescopes, with the PIAACMC design a potentially useful way to reach a narrower inner working angle.  In the mid-infrared, some key optics must operate at cryogenic temperatures in order to minimize thermal background noise, for example a deformable mirror capable of actuation at 5~K.   
 
  \subsection{STARSHADE}
A starshade is an external occulter.  It is an independent spacecraft flying in formation with a space telescope providing a structure that blocks on-axis starlight and directs diffracted starlight away from the telescope with a petal-shaped edge\cite{cash2006}.  The use of a starshade simplifies the design of the space telescope: the picometer-level wavefront errors and stability required for coronagraphy are relaxed to nanometers more typical of a astronomical space telescope, because very little starlight should enter the telescope.   However, a starshade has its own technical challenges, listed in Table~\ref{tbl:TechnologyGaps}.

The basic parameters to consider for a starshade are its size, which is driven by the size of the accompanying space telescope's primary mirror and the wavelength band of interest, and the distance between the starshade and the telescope.  The shape of the petals is optimized for the shadow depth and width, engineering constraints such as petal tip width, and bandpass..  The Exo-S study\cite{exos2015} considered a 30~m diameter starshade co-launched with a 1.1~m space telescope, and a 34~m starshade, launched for a rendezvous with 2.4~m WFIRST.  These starshades would fly in formation at a distance of tens of thousands of km from the telescope.   Most technology development to-date has targeted a 30~m-class starshade.  However, a starshade accompanying a 4~m HabEx or a 15~m LUVOIR will have to be larger (72~m for HabEx and $>$100~m for LUVOIR), requiring different technical solutions in some areas.  A larger starshade would fly further from the telescope to give access to an inner working angle commensurate with the telescope resolution limit (smaller for larger telescopes).

A full-scale pre-launch performance test is impractical because a demonstration of the starshade petal's in-flight starlight diffraction control requires a flight-like Fresnel number $F=r^2/(\lambda Z)$ (where $r$ is the starshade radius, $\lambda$ is the wavelength and $Z$ is the telescope-starshade distance) meaning that the distance in a full-scale test must also be flight-like (30,000 to 50,000~km for Exo-S), which is larger than the Earth's radius.  Therefore designers are forced to rely on subscale testing of the starshade properties to validate precise optical modeling that can be used to create a flight design.  The subscale testing and model effort has led to technology demonstration tests in a tube\cite{leviton2007,samuele2009}, a dry lakebed\cite{smith2016} and on the McMath siderostat\cite{harness2016,novicki2016}.  A sub-scale demonstration at near-flight suppression at similar Fresnel number is underway in a 77~m tube\cite{kim2016} and has demonstrated a contrast within an order of magnitude of in-flight requirement, limited by the subscale mask fabrication.   So far, optical models accurately predict the behavior of the subscale starshades.

Control of light from the Sun scattering off the petal edges drives a requirement\cite{exos2015} that the edge's radius of curvature ($\mu$m) $\times$ reflectivity (\%) $\le$12 $\mu$m$\cdot$\%, independent of the deployment architecture.  Different approaches to obtain the required combination of sharpness and blackness have been investigated using coated\cite{casement2016} or etched\cite{steeves2016} metallic edges.  The petal edge can be manufactured in roughly 1~m sections and attached to the petal structure, making this technology development independent of the overall starshade size.  Optical performance has been successfully demonstrated on small coupon samples, but not yet on 1~m sections.

The starshade concept relies on keeping the starshade structure in precise alignment with the space telescope.  Lateral alignment sensing must be within $\pm$1~m (at 50,000~km, scaling linearly with distance); studies for WFIRST-starshade compatibility have shown this to be feasible with a laser beacon on the starshade and an out-of-band pupil plane camera\cite{martin2015}.   The pupil plane concept is being demonstrated in a subscale laboratory environment.

The final technical challenges involve the mechanical structure of the starshade, which must be stowed to fit within a launch fairing, then autonomously deploy on-orbit to position the petals to better than 1~mm accuracy for a 30~m-class starshade.  The positioning requirements scale nearly linearly with physical size, somewhat relaxing them for larger starshades.  

\begin{figure}
\begin{center}
\includegraphics[width=0.45\columnwidth]{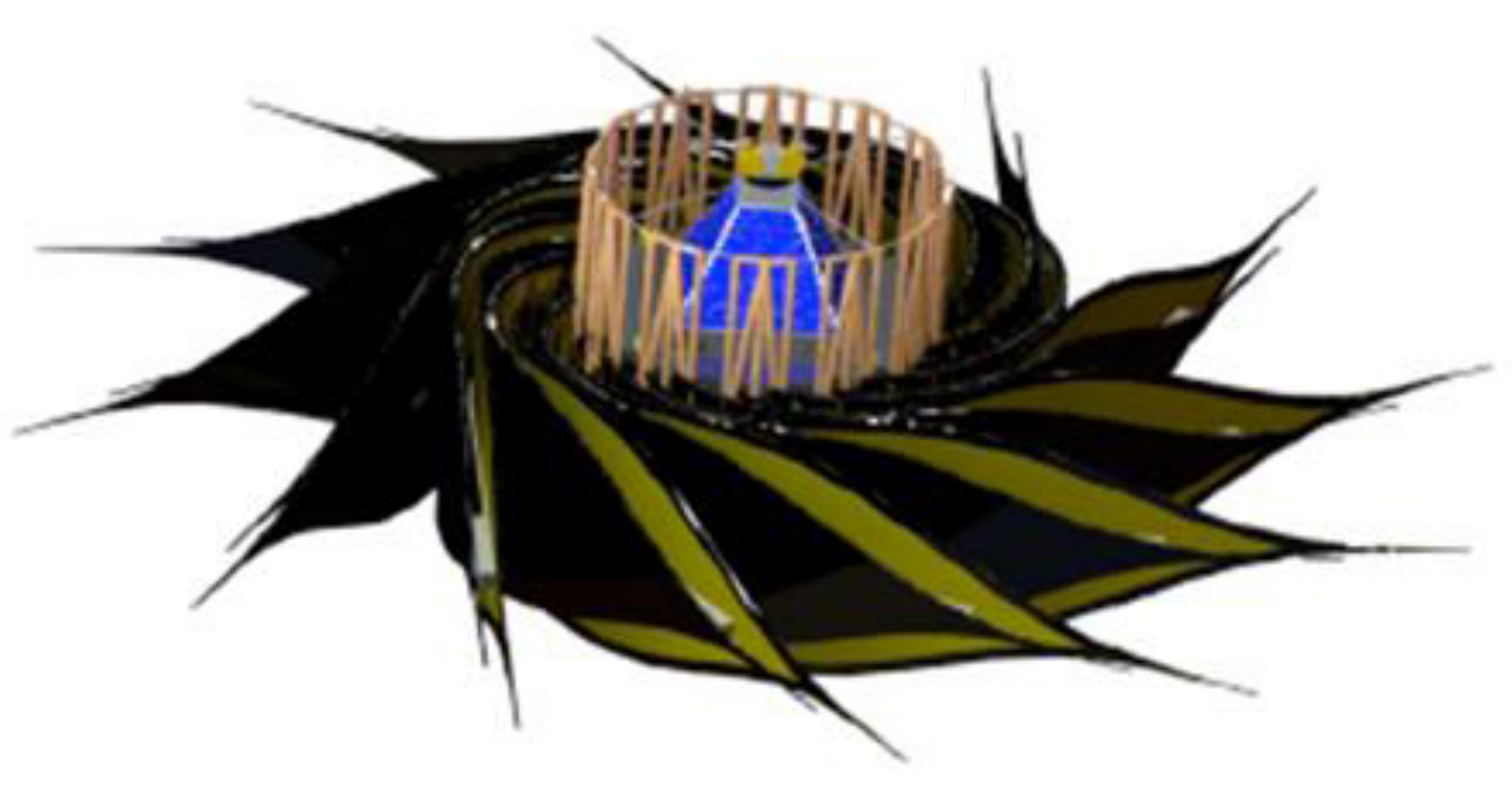} \includegraphics[width=0.35\columnwidth]{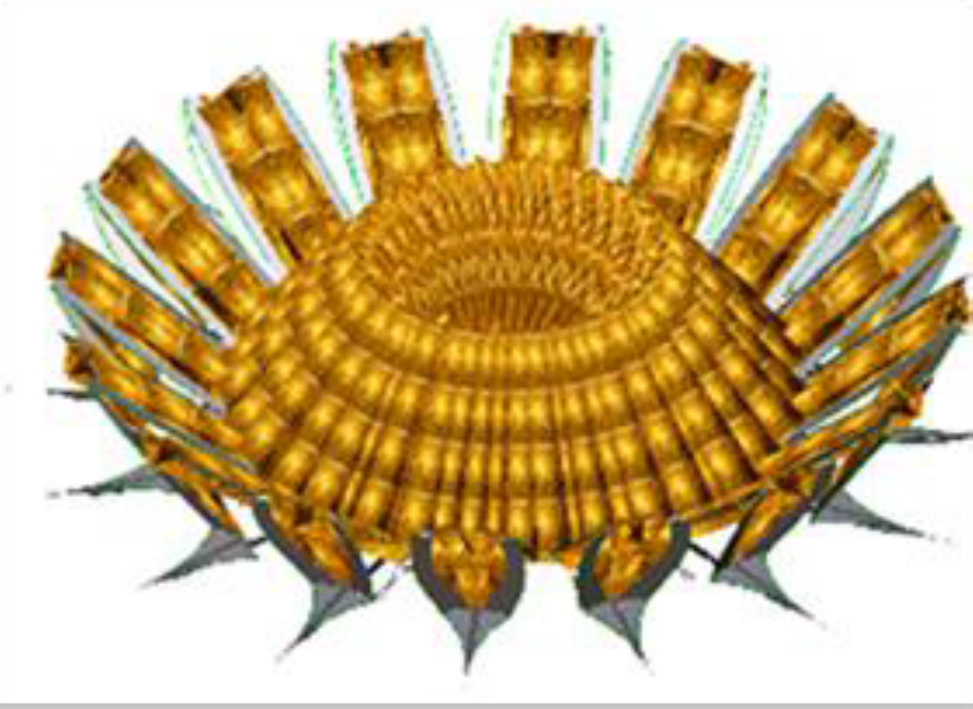}
\caption{\label{fig:starshade_deployment} \textbf{Starshade deployment schemes.}  The right panel shows the ``telescoping boom" approach\cite{lillie2008} and the left panel shows the ``petal unfurling/perimeter truss" method\cite{webb2016} (Image credit: JPL/Caltech (left), NGAS (right).}
\end{center}
\end{figure}

Two design concepts for stowing and deployment are being considered for 20--30~m starshades (Fig.~\ref{fig:starshade_deployment}).  The first, called the petal unfurling / perimeter truss method\cite{webb2016}, which produces the inner circular portion of the starshade using a technique similar to one used for deploying radio frequency antennas.  The petals are manufactured and attached separately, wrapped around the stowed inner disk for launch.  Separate technology development is needed to manufacture starshade petals that can hold shape to sub-millimeter precision after stowage. The biggest petal built to date has been 6~m, which met its structural edge manufacturing envelope of $<$ 100~$\mu$m rms. The inner disk deployment system and the petal manufacturing have been advanced under the TDEM program.  The second approach is a telescoping boom design\cite{lillie2008} that has petals integrated with an inner disk and deploys in an umbrella-like manner.   Work is ongoing to determine whether either of these methods is scalable to a HabEx 72~m starshade, but no work to-date has shown that these are appliciable to a 100+~m starshade.

 NASA chartered the ExEP in 2016 to advance the starshade technologies under a central development activity to accelerate the key technology areas to TRL 5. A technology development plan is expected in Spring 2018 to achieve TRL 5 for all areas by the early 2020s in order to enable a WFIRST-starshade rendezvous mission or probe mission should one be recommended by the 2020 Decadal Survey.
 
  \subsection{SPACE TELESCOPE}
  
  A space observatory for investigating Earth-like exoplanets has several key capabilities: the telescope must be large enough that it can resolve the habitable zones of enough stars that a statistically meaningful survey can be obtained.   Second, in order to use a coronagraph for starlight suppression, the wavefront error from the telescope must be stable to picometers over a detector integration.
 
 The habitable zone of an exo-Earth at 10~pc is 100~milliarcseconds from a Sun-like star (see Fig.~\ref{fig:ContrastPlot}), implying a 3 $\lambda$/D coronagraph needs roughly 25~milliarcseconds angular resolution to detect a planet at 400~nm, or a 3.3~m diameter.  A larger telescope is needed to use the same coronagraph to probe spectral features at longer wavelengths. For example for measuring the 760~nm oxygen line, a 6.3~m telescope is needed.  While technically more difficult,  a  larger primary mirror also improves sensitivity to planets with their greater collecting areas and allows more stars to be searched.   An important, but still not well-known, quantity to consider for telescope size is $\eta_{\rm{Earth}}$, the fraction of Sun-like stars that have Earth-sized planets in their habitable zone.  If $\eta_{\rm{Earth}}$ is 0.1, then a 10~m telescope is needed to find and characterize 30 Earth-like planets\cite{stark2015}.  But if Earth-like planets are more common ($\eta_{\rm{Earth}}$=0.8), then a 4~m telescope can find the same number of exo-Earths.  Further analysis of Kepler data, is expected to give a more accurate value for $\eta_{\rm{Earth}}$ in the near term, which will be an important factor in making the mission-driving choice of primary mirror size.  A report from the Exoplanet Exploration Program Analysis Group's Study Analysis Group 13 will survey this question further\cite{sag_website}.
    
  The largest astronomical space telescope ever flown is Herschel's 3.5~m silicon carbide mirror, which was optimized for longer wavelength (surface figure error of 6~$\mu$m rms). The Hubble Space Telescope, at 2.4~m diameter, is the largest monolith space  astronomical telescope operating in the visible band.  Using segmented mirrors enables even larger primary apertures to fit even larger mirrors into launch fairings and to maintain lightweighting and stiffness.   The launch of JWST, with its 6.5~m  primary mirror consisting of 18 gold-coated beryllium segments, will represent the culmination of many years of technology development that can be leveraged for future space telescopes.
 
As stated in Sect.~\ref{sec:coronagraphs}, a coronagraph with required $10^{-10}$ contrast at a few resolution elements drives the design of a telescope to picometer-class wavefront error stability.  This ambitious stability requirement can be met by designing a telescope and coronagraph with three goals in mind: robustness to disturbance, passive stability, and active control where needed.  Many of the components and subsystems likely to be included in an ultra-stable space observatory, including radius-of-curvature actuation, segment phasing actuation, laser metrology, capacitive edge sensing, thermal sensing, are technologies that exist at various maturity levels, but a systems-level study is recommended to show that all the components can work together successfully.
  
Aside from wavefront stability, several other considerations from coronagraphy enter into design choices for a space observatory.  For managing the diffracted starlight, coronagraph compatibility is best for clear apertures, and in particular off-axis apertures with no obscuring secondary mirror.  Polarization aberration considerations for a coronagraph lead towards a higher f/\# design, though this leads to difficulties in engineering a stable long secondary mirror supports. 
 While not yet explored in detail,  mirror coatings may require improved uniformity to control polarization aberrations.  
  
    \subsubsection{Segmented Primary Mirror}
   Predating the LUVOIR study was the Advanced Technology Large Aperture Space Telescope (ATLAST) concept.  The ATLAST design was studied end-to-end\cite{feinberg2016}, and provides an example of a 9~m segmented mirror with excellent wavefront stability that can be scaled to a 15~m LUVOIR.   The mirror segments themselves are 1.5~m-class mirrors which are relatively easy to construct, lightweight, and handle compared to large monoliths.  Glass (ULE and Zerodur) and silicon carbide (SiC) with a nano-laminate have achieved ~10 -- 15~nm surface figure error (SFE) at high spatial frequencies on 1.5 m-class segments. Achieving $<$ 5~nm SFE for glass and cladded SiC is likely to be straightforward, but expensive, by applying additional iterative polishing steps. Non-NASA work with silicon-carbde and nano-laminates have produced engineering-level prototype mirror systems and demonstrated them in a relevant environment.    Furthermore, it is not clear if glass segments will have the surface figure control authority required to achieve the achieve the contrast requirements and if deformable mirrors further downstream can do the job.  The LUVOIR and HabEx design teams are studying this.
   
 Aside from the construction of the mirror segments themselves, the mounting and (likely actively actuated) backplane technology must be able to maintain a stable primary mirror shape to picometer precision and active correction at a variety of time scales is likely to be needed to correct thermal and vibrational disturbances.  Actuation of mirrors has been demonstrated, but sensing at the picometer level metrology will be challenging.  Capacitive edge sensors, like those used on current and planned ground-based observatories such as Keck and TMT, can be read out at kHz bandwidth, but must be improved to the picometer level of sensitivity.    Laser metrology has been advanced to achieve picometer-scale measurements, but only at refresh rates corresponding to slow thermal drift time scales.
   
   Vibration isolation to separate sources of vibration such as reaction wheels from the mirror segments is likely to be needed for a segmented space telescope utilizing a coronagraph.  A contact-free isolation stage called the Disturbance-Free Payload, designed by Lockheed Martin as a candidate for JWST \cite{pedreiro2012}, and has shown to provide sufficient isolation for the ATLAST 9~m concept.
   
Low-order wavefront sense/control inside the coronagraph can potentially correct many disturbances, and the WFIRST LOWFS/C system can sense Zernike modes up to Z11.  A space telescope with a segmented primary mirror could in principle use a pupil-plane segmented deformable mirror to correct relative piston, tip and tilt of mirror segments, though this concept has not yet been demonstrated.

A full investigation of these trades giving requirements on individual subsystems for a large segmented space mirror compatible with an exo-Earth-characterizing coronagraph will come from the LUVOIR concept study.

  \subsubsection{Monolith Primary Mirror}
  The HabEx study is investigating 4~m-class monolith mirrors as a telescope solution that works well with a coronagraph. While there are fewer degrees of freedom for a monolith space mirror than for a segmented mirror, the picometer-level requirements on wavefront errors are still extremely difficult to achieve.  While 8~m and larger high-quality monolith mirrors with 7 -- 8~nm surface-figure error exist for ground-based applications (for example Subaru, VLT), space applications traditionally require lower areal density and stiffness (expressed as the first resonant mode of the mirror).    The wavefront stability, affecting coronagraph performance, scales with the square of the first resonant frequency mode, linearly with the mass, and linearly with the disturbance acceleration.    

  A lightweighted, and stiff, closed-back ULE glass primary mirror suitable for up to 4~m monolith mirrors was investigated with the ATMD-2 project \cite{egerman2015} though manufacturing difficulties were found.   The planned Space Launch System (SLS) rockets will allow the launch of more massive mirrors, potentially reducing complications of lightweighting, though open-backed mirrors will be less stiff.  Thermal control to roughly the 1~mK level is also needed for large monoliths given the CTE uniformity of typical mirror glass\cite{jedamzik2016}.  Additional mass has some advantages in reducing dynamic wavefront errors from disturbances and slower response to thermal transients.  Using microthrusters rather than reaction wheels for pointing the observatory (following GAIA),  could additionally reduce the disturbances to the system.   A clear aperture allows several coronagraph architectures to be considered, including a high-charge vortex coronagraph, which has nearly zero sensitivity to several Zernike wavefront error modes, easing some wavefront error stability requirements. A more massive and less stiff mirror also pushes the limits of the ability to measure and remove the effects gravity sag to determine that the on-orbit surface figure meets requirements.   A full study showing the requirements on individual subsystems and the feasibility of a 4~m space monolith mirror is forthcoming from the HabEx concept study.

  \subsection{DETECTORS}
  
Rauscher et al. (2016)\cite{Rauscher2016}  recently reviewed the state of visible-band and near-infrared detectors and summarized improvements needed for a biosignature search, in addition to suggesting investments towards maturing these technologies.    For exoplanet direct imaging, given the low flux targets (Absolute V-band magnitudes $\sim$30), high quantum efficiency, low noise (including dark current),  moderate spectral resolution (R=100), and large format arrays are typical requirements.  

Passively cooled photoconductors are relatively mature, with recent space mission heritage, and are already close to the needs of an exo-Earth mission in the visible and near-IR bands.  Leading candidates include electron-multiplying charge-coupled devices (EMCCD)  shortwards of 1~$\mu$m and HgCdTe arrays or HgCdTe avalanche photodiode (APD) arrays in the near-infrared ($\lambda >$0.4~$\mu$m).

The 1k$\times$1k EMCCD arrays manufactured by e2v and selected for the WFIRST/CGI visible band detector\cite{harding2015} have met WFIRST/CGI technology milestones, showing dark current of 7$\times$10$^{-4}$ e$^{-}$/px/s, clock-induced charge of 2.3$\times$10$^{-3}$ e$^{-}$/px/frame, effective read noise $<$0.2 e$^{-}$ rms, after irradiation when cooled to 165~K.  This meets the needs of WFIRST, but improvement in radiation hardness is desirable for a longer mission.  Straightforward changes in the manufacturing of the WFIRST detectors are likely to make incremental improvements in the near term.  Longer term, p-channel semiconductors, for example Hole-multiplying CCDs\cite{rauscher2016_npr}, are likely to be less radiation-sensitive.  Skipper CCDs, developed for dark matter direct detection, use a readout scheme that gives sub-electron read noise without avalanche gain\cite{tiffenberg2017}, but have unknown space radiation tolerance.  CMOS detectors offers great potential for radiation hardness and low read noise, but has little heritage for low-light astronomical applications.

In the near-infrared, HgCdTe arrays from Teledyne  have been advanced for use by JWST\cite{rauscher2014}, Euclid\cite{Waczynski2016},  and the WFIRST wide-field instrument \cite{piquette2014} with good sensitivity to wavelengths as long as 5~$\mu$m.  These are radiation-hard, and have been shown to reach dark current levels of $<$10$^{-3}$~e$^{-}$/px/s.  Read noise is still a challenge: 5~e$^{-}$ rms/px has been achieved, while $<$1e$^{-}$ rms/px is needed.  Some improvements can be made using readout schemes that effectively allow averaging over the read noise.  Additionally, investigation into the sources of noise internal to the readout electronics can potentially yield improvements.

HgCdTe APDs, manufactured by Leonardo/Selex, have shown great promise in effective read noise, by using a gain factor of roughly 10, but still suffer from unacceptably high dark current (well above 1~e$^{-}$/px/s).  The high dark current is attributed to ROIC glow and is expected to be reduced greatly in the short term, with help from NASA APRA funds.

If a future exo-Earth-finding mission includes active cryogenic cooling, single-photon detectors based on superconducting thin films offer a number of advantages, considering the very low flux from exo-Earths: namely the noiseless detection of photons.    Microwave Kinetic Inductance Detectors (MKID) measure the arrival of single photons with a shift in kinetic inductance of a superconductor.    MKIDs were invented and developed mainly for use in the millimeter-wave band, but recent work has begun to optimize these for ground-based visible-band astronomical applications, in part with funding from the NASA APRA program. Read in a resonant circuit at very low temperature (typically 100~mK), these have effectively zero noise, though so far only limited energy resolution in the visible band (R=10)\cite{mazin2013}.   Transition Edge Sensor (TES) microcalorimeters measure the energy of a photon via the temperature rise of an absorber.  TES detectors have been advanced by work in sub-mm wavelengths as well as in the X-ray band, but need optimization for the visible and near-IR bands.  Additionally, vibrations from a mechanical cryocooler can create difficulties for a coronagraph's need for extreme wavefront stability. The MIRI instrument on JWST includes a Joule-Thomson cooler that would create a 3~nm rms wavefront error\cite{Rauscher2016},  larger than a high-contrast coronagraph requires.  Hence, the HabEx and LUVOIR studies have not included active cooling in their trade space.  If the advantages of superconducting detectors are considered to be essential,  the disturbance sources from active cryogenic coolers would need to be reduced by 2 -- 3 orders of magnitude for a coronagraph mission.  A starshade-only mission could potentially take advantage of actively cooled detectors due to less stringent stability requirements on the space telescope.

  \subsection{GROUND SUPPORT}
  
  Ground-based measurements can provide valuable information for space-based searches for exo-Earths.   NASA has invested in the Hunt for Observable Signatures of Terrestrial Systems (HOSTS) survey\cite{danchi2016} with the Large Binocular Telescope Interferometer to determine the typical amount of exo-zodiacal dust around nearby stars.  This survey, expected to complete in 2018, has a direct impact on technology needs: if exo-zodiacal dust is typically very dense, direct imaging science goals could require even larger space telescopes to reduce this source of noise.
  
  Additionally, precision radial velocity (RV) techniques measure the Doppler shift in stellar absorption lines as orbiting planets cause a gravitational recoil in the star.  RV has been a historically important technique for discovering exoplanets, and recently was used to discover a planet orbiting our closest neighboring star, Proxima Centauri\cite{angladaescude2016}.    RV obtains the crucial planetary mass measurement that is inaccessible to direct imaging and transit photometry measurements, allowing a planet's density and atmospheric scale height to be inferred.  In addition, dedicated ground-based RV surveys in the Northern and Southern hemispheres could improve the science return of a space mission.  By pre-selecting stars with known planets, and using RV-measured planet orbits to schedule observations for maximum brightness, the yield of a space mission will be improved.  
 
  The reflex motion of a Solar-mass star due to an orbiting Earth-mass planet at 1~AU is $\sim$10~cm/s over 1 year.  Detection of the Doppler shift of spectral lines to this precision requires control of instrumental sensitivity and removal of systematics to better than this level on the relevant time scales, or nearly an order of magnitude improvement in the current-state-of-the-art\cite{fischer2016}.   The development of the NEID instrument for the WIYN telescope\cite{schwab2016} under the ExEP's NN-EXPLORE program will take an important step in this direction, with a goal of 27~cm/s precision, but further understanding of stellar noise and telluric atmospheric effects, as well as additional improvement in instrumental stability will be needed beyond NEID.     Additional sensitivity may be gained by including an RV instrument on a space telescope, and to investigate this possibility, an Astrophysics Probe study was funded by the NASA Astrophysics Directorate in 2017 to study the improvement in measurement errors that could be obtained from a space platform.  Observing form space bypasses the microtelluric absorption lines from the Earth's atmosphere, a major systematic error faced by ground-based RV measurements.    RV measurements in space would also open near-IR wavelengths to sensitive measurement which might help in reducing the effects of stellar noise.  Space-based RV techniques, or ground-based RV supporting a space mission, may require new technology to achieve its sensitivity needs.
  
  \section{THE PATH AHEAD}
  \label{sec:path_forward}

\begin{figure}
\begin{center}
\includegraphics[width=\columnwidth]{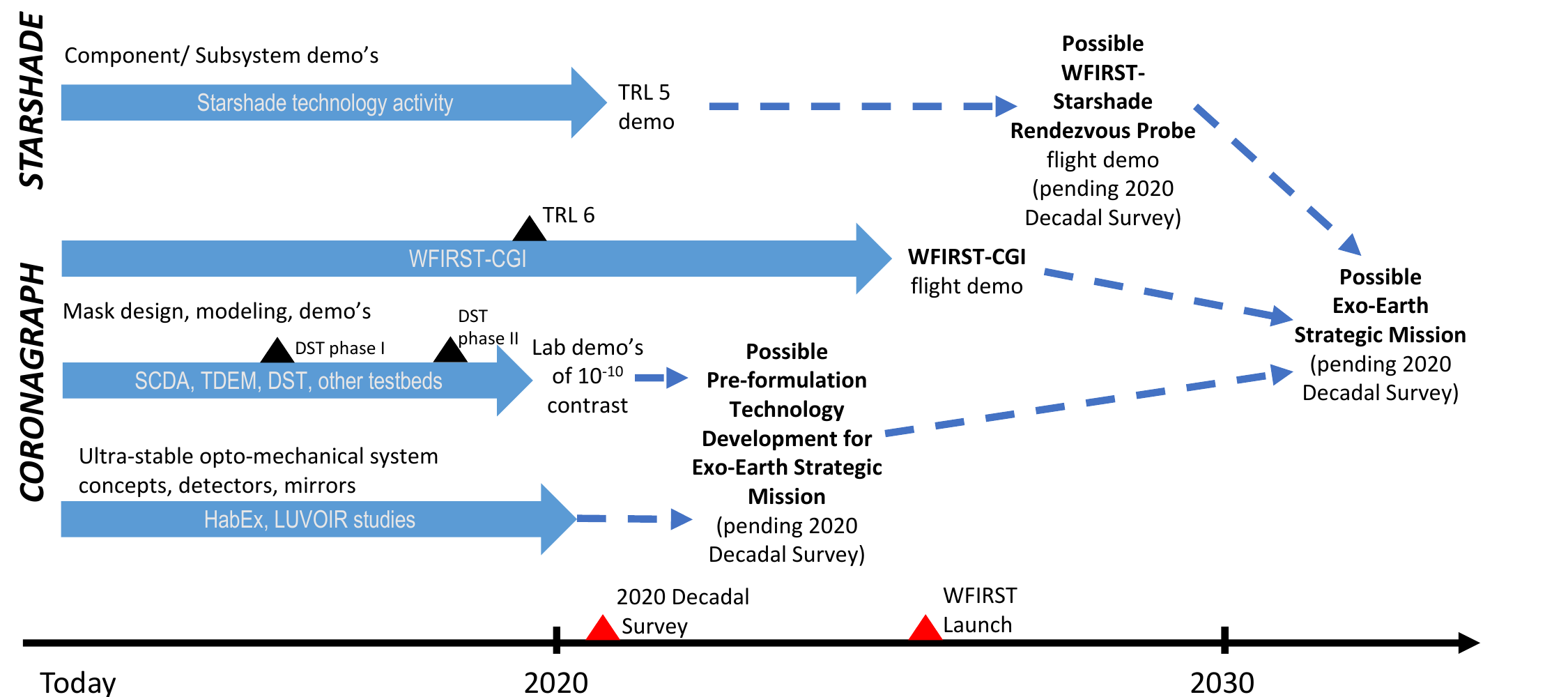}
\caption{\label{fig:RoadMap} \textbf{A possible technology maturation path towards a strategic exo-Earth-finding mission.} Solid blue arrows represent existing efforts and dashed arrows are efforts pending the 2020 Decadal Survey's recommendations.  Current efforts leading towards the strategic mission include the starshade technology activity, WFIRST Coronagraph Instrument (CGI), the Segmented Coronagraph Design Analysis (SCDA) study, the Technology Development for Exoplanet Missions (TDEM) portion of the Strategic Astrophysics Technology (SAT) program, the Decadal Survey Testbed (DST), and the Habitable Exoplanet Imaging (HabEx) Mission, and Large Ultraviolet-Optical-Infrared Surveyor (LUVOIR) mission studies.}
\end{center}
\end{figure}

NASA's implementation of the New Worlds Technology Development Program\cite{AIP2016} recommended by the 2010 Decadal Survey has helped to mature exoplanet direct imaging technology, and will continue to do so in the near term.  This is being accomplished through a competed TDEM program, upgrading the High Contrast Imaging Testbed for coronagraph demonstrations, the ExEP starshade technology development activity, along with the Decadal Mission concept studies.  Figure~\ref{fig:RoadMap} shows a possible technology maturation path to advance starlight suppression and other technologies from the present day to an exo-Earth-characterization mission presumed for the 2030's, pending the 2020 Decadal Survey recommendation.  Table~\ref{tbl:TRLadvancement} shows some of the ongoing activities maturing coronagraph, starshade, wavefront sensing and control, and detector technology.   The first column shows progress that is likely to be made by 2020 with existing efforts.  The middle column shows the level that will be reached after WFIRST, and the last column attempts to capture remaining steps that need to occur to be ready for an exo-Earth-finder.  This table is likely incomplete but attempts to capture the major areas of on-going, planned, and possible activities.

The WFIRST mission, with the CGI technology demonstration instrument representing the first high-contrast direct imaging system in space, will play an essential role in continuing to mature coronagraph, wavefront sensing, deformable mirrors, and detector technology.  The HabEx and LUVOIR studies will continue to develop their spacecraft and instrument designs, providing exoplanet-imaging reference missions.   The studies have already played an important role, creating more detailed requirements and a better understanding of the technology needs. 

The longer-term path for exoplanet direct imaging technology development will depend largely on the recommendations of the 2020 Decadal Survey.  In particular, the committee may or may not, endorse an exo-Earth-finding mission and define particular objectives.  HabEx and LUVOIR will be presented as two potential exoplanet-driven options to be considered.  Additionally, the 2020 Decadal Survey may, or may not recommend that NASA proceed with a starshade rendezvous mission with WFIRST. 

\subsection{SAT}
NASA's SAT program, managed out of the Astrophysics Division, is a competed grant opportunity\cite{perez2014} with the goal of maturing technology to the point that it can be used in a flight mission.  Typically this involves the development of technology from a laboratory proof-of-concept (TRL 3) to a medium-fidelity technology tested in a relevant environment (TRL 5).    The ExEP facilitates the Technology Development for Exoplanet Missions (TDEM) portion of  SAT, and the Program's prioritized technology gap list informs the specific language of the TDEM call for proposals, ensuring that the proposals are oriented towards enabling NASA's next exoplanet missions.  The Cosmic Origins (COR) Program portion of SAT\cite{pham2016} has traditionally included in its purview technology for optical/infrared/UV space observatories, including detectors and mirrors.  It therefore has some overlap with TDEM, though often exoplanet science requirements are different, and often more stringent, than those needed for general astrophysics.    COR and ExE Program technologists ensure that technology needs are all addressed by at least one of the programs, in part through cross-representation on the review boards of the annual selection and prioritization of technology gaps.
 
 The TDEM program has funded the advancement of much of the existing space-based starlight suppression technology, covering multiple coronagraph and starshade elements.  The results from the completed demonstrations are captured in TDEM Milestone Final Reports, and are available on the ExEP website\cite{tdem_website}.   Active TDEM awards are advancing vortex coronagraph masks expected to demonstrate high contrast in a broad band in the HCIT  (TDEM-14, PI Serabyn) and the understanding and mitigation of polarization aberrations in coronagraphs (TDEM-15, PI Breckinridge).  Visible Nuller Coronagraph (VNC) technology is advancing through a TDEM-13 (PI Hicks)\cite{hicks2016}. A TDEM-14 (PI Bendek) is investigating a concept combining precision astrometry and a Phase Induced Amplitude Apodization (PIAA) coronagraph to be able to determine the planet mass as well as performing spectroscopy.  TDEM awards aimed at advancing deformable mirror technology through environmental testing will continue (PI Helmbrecht TDEM-10 and PI Bierden TDEM-10). 

The ExEPÕs technology gap list captures those technologies that enable and enhance the direct imaging and characterization of exoplanets. It is updated annually through the ExEP Technology Selection and Prioritization Process which includes inputs from the community and is reviewed by an outside committee.    This process includes input from the Decadal Survey mission studies, mainly from HabEx and LUVOIR, as well as from OST.  The gap list is prioritized\cite{TechnologyPlanAppendix2017} to make it more useful in guiding the TDEM call for proposals; starlight suppression technologies have been prioritized highly as they are deemed enabling technologies.
 
The Decadal Survey Tested (DST) is an upgrade to a testbed within the HCIT begun in 2017 that will enable vacuum demonstrations of 10$^{-10}$ contrast on both clear and segmented/obscured apertures.  In commissioning the DST, an initial demonstration will attempt to achieve 10$^{-10}$ broadband with a clear aperture.  The second phase, planned for 2019, will demonstrate the 10$^{-10}$ contrast with a segmented pupil mask, forcing a coronagraph to handle the additional diffraction from the segment edges and central obscuration. Phase III will include a segmented telescope simulator introducing simulated phasing errors between segments and other wavefront disturbances.  In the coming years, with access to the DST, TDEM-funded starlight suppression demonstrations could reach the 10$^{-10}$ contrast goal by the time of the 2020 Decadal Survey.
 
   \subsection{WFIRST}
   The CGI, with its active wavefront control subsystem, will be the first high-contrast coronagraph in space, making a large step across the contrast performance gap (Fig.~\ref{fig:ContrastPlot}).  In 2017, CGI met its last technology milestone and achieved TRL 5.  This work has pushed the state-of-the-art in Shaped Pupil and Hybrid Lyot coronagraph technology, working with WFIRST's highly obscured pupil.  Additionally it has advanced LOWFS/C and deformable mirrors validating the optical performance models in its relevant environments.  WFIRST-CGI has flight-qualified an e2v EMCCD array, testing its optical performance after simulated exposure to the L2 space radiation environment.  The PIAACMC technology has also been advanced as a backup technology, but fell short in performance.  As the WFIRST project transitions to a flight design and build phase,  some parts of the CGI design will be frozen and managed through configuration control, but in important areas developments will continue.   The EMCCD detector will be further rad-hardened, and investigation into the performance of the deformable mirrors will continue.   Even longer term, for the next 3+ years, the coronagraphic masks can continue to be improved and swapped in CGI's filter wheel.  Additionally, CGI is required to be re-programmable on orbit, allowing improved LOWFS/C algorithms and high-order wavefront control algorithms to be implemented and tested all the way through on-orbit operations.    

\subsection{Starshade-WFIRST Rendezvous}
A starshade rendezvous late in the WFIRST prime mission, if recommended by the 2020 Decadal Survey, is likely to be the fastest way to image an Earth-like exoplanet in the habitable zone of a nearby Sun-like star.  While a starshade-WFIRST rendezvous mission has not been initiated by NASA, there are ongoing developments in advancing starshade technology.  First, the WFIRST project has been directed to maintain its study of starshade compatibility.  In particular, this will help to advance the formation flying sensing technology.  The Astrophysics Probe selection in 2017 included funding for an update of the WFIRST-starshade rendezvous concept (PI Sara Seager).

Secondly, technology development funds from the ExEP are advancing the maturity of starshade technology.   The starshade technology development activity is currently developing a plan to mature starshade hardware to medium-level fidelity (TRL~5) by the early 2020's.  The most challenging areas for advancement are the mechanical technology involving precision deployment of a 34~m structure.

A future exo-Earth-finding mission is likely to have a larger ($\ge$4~m) primary mirror, meaning that an accompanying starshade is likely to be much larger than one optimized for a WFIRST rendezvous.  Nonetheless, the successful operation of a 34~m starshade would demonstrate the technology in an operational environment, and provide essential lessons and risk mitigations for the future exo-Earth finding mission.

\subsection{HabEx}
HabEx and LUVOIR can be considered to be two design concepts on a continuum of exoplanet imaging and characterization capabilities. The mission studies are both taking systems-level views of a space mission for discovery of exo-Earths and the spectral search for biosignatures.  Their designs will help set requirements on subsystems and components, allowing trades to be understood and giving a better sense of what technologies are likely to require the most development.  The final reports for these studies are due to be complete in spring of 2019 and submitted to the Decadal Survey panel shortly thereafter.  

The first baseline concept for HabEx is an off-axis telescope with a 4~m monolith primary, making it the largest monolith to ever fly in space, and using both a coronagraph and a starshade for starlight suppression.    As stated earlier, the 10$^{-10}$ contrast requirements of a coronagraph maps into a wavefront stability error requirement better than 1~nm rms.  The clear (not obscured by a secondary mirror and its supports) and unsegmented aperture is considered ideal for a coronagraph and allows the use of a Vortex Coronagraph or a Hybrid Lyot coronagraph for passive stability to some disturbances.  The least mature areas of technology development for HabEx are related to the starshade;  the mission is likely to need a 72~m-diameter starshade which is roughly twice as large as the starshade proposed for a WFIRST rendezvous mission.  

HabEx is studying the use of a 4~m Zerodur glass monolith mirror for compatibility with a 10$^{-10}$ coronagraph.  It must be thermally controlled and resilient to mechanical disturbances to the required wavefront error.   The study will consider reduction of the mechanical disturbance environment through the use of microthrusters for fine pointing control.    A telescope with a 6.5~m segmented primary mirror will also be considered as a second architecture by the HabEx team later in the study, and is likely to have a different set of technology challenges.

\subsection{LUVOIR}

The first baseline telescope concept considered by LUVOIR is an on-axis 15~m primary mirror, constructed from 120 $\sim$1.5~m segments and using a coronagraph for starlight suppression.

The two main areas of technology development for the LUVOIR concept are 1) the ultra-stable opto-mechanical structure required to form a large primary mirror out of segments, with the required picometer-class wavefront error stability, and 2)  segmented-aperture coronagraphy at the 10$^{-10}$ contrast level\cite{bolcar2016}.  For the former challenge, the team is building on experience gained from the ATLAST 9~m telescope study\cite{feinberg2016} which had a similar architecture to designs under consideration for LUVOIR.  The LUVOIR team is performing model-based analysis to show that the wavefront error requirement can be met, and to broadly flow requirements into different subsystems with multiple levels of open and closed-loop control.  Laser metrology, high-bandwidth capacitive edge sensing, thermal control, dynamic isolation systems,  and low-order wavefront sensing and control are all subsystems that may be used to achieve the overall stability.

To meet the challenge of segmented aperture coronagraphy, the ExEP's SCDA study will continue to advance design for coronagraphs that work with segmented apertures through September of 2018\cite{scda_website}.  In addition to the baseline SCDA set of segmented and obscured apertures, the SCDA teams are working with the LUVOIR team to include the LUVOIR aperture in their studies.   As stated in Sect.~\ref{sec:coronagraphs}, the Apodized Pupil Lyot Coronagraph (APLC) approach has achieved the most success to-date.  Robustness calculations (effects of optical misalignments) performed by the SCDA will also provide input for LUVOIR's design choices.  As of this writing, the APLC was the only coronagraph design meeting LUVOIR requirements; response to segment-to-segment phasing errors is under study.

LUVOIR also plans to consider an architecture based on a 9~m segmented telescope, though this architecture has not yet been addressed in detail.    LUVOIR may additionally consider using a starshade as a future rendezvous option, though with a large primary mirror, a starshade could potentially reach sizes of order 100~m or larger in diameter depending on trades involving the needed inner working angle and wavelength coverage.  Deploying a large starshade autonomously is a challenge and on-orbit robotic assembly is an approach that may be required, but is not part of the LUVOIR study.

 \subsection{Decadal Survey}

While the technology development described in this paper is aimed towards an exo-Earth-finding mission for the 2030's, the path forward will depend on the recommendations of the 2020 Decadal Survey from the National Academies.  The 2020 Decadal Survey committee will begin meeting in 2018 and is expected to release its report in December 2020.  The report will make a recommendation for the nation's astrophysics priorities, including which large NASA mission(s) should  follow JWST and WFIRST, and whether a starshade mission should be developed to rendezvous with WFIRST at the end of its prime mission.

In the time leading up to the Decadal Survey committee meetings, many technologies will become more mature thanks to ongoing efforts in a wide variety of areas, including in the WFIRST-CGI project (Table~\ref{tbl:TRLadvancement}).  These advancements could help the committee decide on the scope of a potential exoplanet mission.  Once the Decadal Survey committee makes its recommendations, the technology development path could potentially be altered.   Similarly, should the Decadal Survey panel recommend a large mission to search for and characterize exo-Earths, it may or may not look like HabEx or LUVOIR, possibly adjusting the requirements described in this paper.

\scriptsize
 \begin{center}
 \rowcolors{2}{white}{gray!10}
 \begin{longtable}{K{2.8cm}K{3.9cm} K{3.9cm} K{3.9cm}}
 \hline
 Technology & 2017 -- 2020 & After WFIRST&Remaining needs\\
 \hline
 \endhead
 Ultra-stable Opto-mechanical System with Large Aperture Monolith Mirror&
 \begin{itemize}
 \item Advancement of ULE lightweighting (SAT/COR)
 \item Studies of Zerodur 4~m monolith (HabEx)
 \end{itemize}
& 
&
\begin{itemize}
\item Demonstrate at sub-scale
\end{itemize}
\\
  
Ultra-stable Opto-mechanical System with Large Aperture Segmented Mirror&  
 \begin{itemize}
 \item System level studies (LUVOIR)
 \item Future large-aperture segmented space telescope study (NASA Cosmic Origins Program)
 \end{itemize}
 &
 & 
 \begin{itemize}
 \item Perform hardware demonstrations of components and sub-systems meeting performance requirements.
 \item Demonstrate fully integrated mirror segments.
 \item Demonstrate closed-loop sense/control architecture between 2 or more subscale segments.
 \end{itemize}
\\
 
 Clear-aperture Coronagraph & 
\begin{itemize}
\item Coronagraph design (HabEx)
\item Charge-6 and charge-8 Vortex masks demonstrated (TDEM-14)
\item DST phase I (ExEP)
\item Possible coronagraph mask demonstrations (NASA ROSES SAT/TDEM)
\end{itemize}
 &
 & 
 \begin{itemize}
 \item Dynamic testing of coronagraph with monolith telescope simulator
 \end{itemize}
 \\
 
 Segmented-aperture Coronagraph & 
\begin{itemize}
\item SCDA (ExEP)
\item Coronagraph design (LUVOIR)
\item Visible Nuller Coronagraph (TDEM-13)\cite{hicks2016}
\item DST phase II (ExEP)
\item APLC testbed (STSci)
\item Segmented coronagraph testbed (Caltech)
\end{itemize}
 &
 \begin{itemize}
 \item Flight qualified DMs, LOWFS/C, HLC and SPC coronagraphs with WFIRST obscured pupil.
 \end{itemize}
 & 
 \begin{itemize}
 \item Demonstrate high actuator count and (if needed) higher stroke DMs
 \end{itemize}
 \\
 
 Ultra-low Noise Near-infrared Detectors & 
 \begin{itemize}
 \item HgCdTe APD dark current improved (APRA)
  \item improvement of MKIDs (APRA)
 \end{itemize} & 
 &
 \begin{itemize}
 \item HgCdTe read noise improved
 \end{itemize}\\
 
 Ultra-low Noise Visible-light Detectors & 
 \begin{itemize}
 \item EMCCD rad hardness improved (WFIRST)
 \item improvement of MKIDs (APRA)
 \end{itemize}
 &
 Gap could be closed if energy-resolving detectors not needed. 
 &
\\
 
 Starshade Diffracted Starlight Suppression & 
 \begin{itemize}
 \item Subscale demonstration of flight starlight suppression at flight Fresnel number (Starshade Tech. Activity)
 \end{itemize}
& 
If starshade rendezvous occurs, this gap is closed.
 &
\\ 
 
 Starshade Scattered Sunlight Suppression & 
 \begin{itemize}
 \item Meter-class optical edges manufactured that meet optical and mechanical requirements (Starshade Tech. Activity, SBIR)
 \end{itemize}
  &
  If starshade rendezvous occurs, this gap is closed.
  & 
\\ 
  
 Starshade Lateral Formation Sensing &
 \begin{itemize}
 \item Demonstration of benchtop sensing concept closing a control loop (WFIRST, Starshade Tech. Activity)
 \end{itemize}
  &
If starshade rendezvous occurs, this gap is closed.  
  &
\\ 
   
 Starshade Petal Positioning & 
 \begin{itemize}
 \item Study 70~m-class starshade (HabEx)
 \item Perimeter truss deployment motor (SBIR)
 \item Down-select petal deployment architecture (Starshade Tech. Activity)
 \end{itemize}
 & 
 \begin{itemize}
 \item If starshade rendezvous occurs, on-orbit demonstration of a $\sim$30~m starshade deployment, scalable to roughly 70~m-class starshade
 \end{itemize}  
 &
  \begin{itemize}
 \item 100~m+ class starshades possibly need redesign.
 \end{itemize}  
\\ 

Starshade Petal Shape and Stability & 
\begin{itemize}
 \item Down-select petal deployment architecture (Starshade Tech. Activity)
\end{itemize}
 &
  \begin{itemize}
 \item If starshade rendezvous occurs, on-orbit demonstration of a $\sim$6~m petal, acting as a half-scale model for a roughly 70~m-class starshade.
 \end{itemize}  
 &
\begin{itemize}
 \item demonstrate $\sim$half-scale petals for 100~m+ class starshades.
 \end{itemize}   
\\ 
 
 \hline
 \rowcolor{white}
 \caption{\label{tbl:TRLadvancement} \textbf{Advancement of exoplanet direct imaging and characterization technology towards infusion into a flight mission}.  The columns show steps expected to be made prior to 2020 with current investments, and steps that will be taken once WFIRST is successful, and finally, remaining steps needed.  Due to the difficulty of creating a complete list of ongoing work, this list is likely incomplete and we invite the reader to contact the authors to add any entries missing from the Table.  Updates will be added to the ExEP Technology Plan Appendix\cite{TechnologyPlanAppendix2017}.}

 \end{longtable}
 \end{center}
\normalsize
\subsection{Longer Term Vision}
In 2013, NASA released \textit{Enduring Quests, Daring Visions: NASA Astrophysics in the Next Three Decades}\cite{30year2013}, presenting a long term vision for NASA astrophysics.  Of the three major scientific questions addressed in this plan,  "Are We Alone?" addresses the search for life in the Universe studying exoplanets.  The report looks beyond an initial exo-Earth direct imaging and characterization mission such as LUVOIR or HabEx, and envisions an Exo-Earth Mapper, capable of resolving surface features of distant Earth-sized planets.  Exo-Earth Mapper would allow a definitive search for, and study of, life on other worlds by enabling the study of clouds, oceans, continents, and land cover on an exo-Earth.  Such a mission involves new technology well beyond the scope of what is currently under development, most likely using space-based interferometry to obtain the necessary angular resolution and light-collecting area for these ambitious goals.  Future Decadal Surveys will establish whether developing technologies towards these scientific goals are a priority.

  \section{CONCLUSION}
  \label{sec:conclusion}
  
  There are many technological challenges to achieve the ambitious goal of imaging an exo-Earth and looking for spectroscopic evidence of life in its atmosphere. The key technology to directly imaging Earth-size exoplanets in the next couple of decades is starlight suppression with a starshade or a coronagraph.   NASA and the exoplanet community do not know enough at this time to decide which of the two approaches is most likely to be successful and are advancing both technologies until more is learned.   In addition to starlight suppression with a starshade or coronagraph, advancement is needed in telescope (segmented and monolith primary), telescope stability, and visible-band and near-infrared detector technologies.  There are a number of advancements occurring now, funded by NASA and other sources.  Ongoing work is also determining other missing capabilities through the development of systems-level studies and mission concepts.

The 2020 Decadal Survey will establish the nation's astrophysics priorities for the future, and depending on the Survey's recommendation, the technology needs for direct imaging may need to be updated or prioritized differently.    As an example, to interpret the spectroscopic measurements of an exoplanet, the mass must be measured.  Mass measurements via precision radial velocity or astrometry may need technology development in other areas.   

As direct imaging technologies described here become ready for infusion into a mission whose goal is to find Earth-like exoplanets in the habitable zone of their stars, a major step will have been taken towards the search for life in the Universe.   The ongoing HabEx, LUVOIR, and OST design studies are examining specific mission concepts.  The NASA astrophysics 30-year roadmap\cite{30year2013} presents a possible vision towards NASA's long-term goal of resolved images of Earth-sized exoplanets revealing continent-scale features. The development of new technology for exoplanet missions will continue towards these even more ambitious goals.

  \acknowledgments
 This work was carried out at the Jet Propulsion Laboratory, California Institute of Technology, under a contract with the National Aeronautics and Space Administration.   We thank Eric Mamajek, John Gagosian, Steve Warwick, Ilya Poberezhskiy, Chas Beichman, Keith Warfield, R\'emi Soummer, and Stuart Shaklan for helpful comments.
  
\bibliographystyle{spiebib}
\bibliography{spie_roadmap}
 
 \end{document}